\documentclass[aps,preprint,nofootinbib,preprintnumbers]{revtex4}
\usepackage{graphicx,amsmath,amssymb,bm,subfigure,comment}
\usepackage[english]{babel}
 \usepackage{color}
\usepackage{slashed}
\usepackage{amsmath}
\usepackage{wasysym}
\usepackage{latexsym}
\usepackage{graphicx}
\usepackage{amssymb}
\usepackage{amsmath}
\usepackage{mathrsfs}
\usepackage{fancyhdr}
\usepackage{color}

\newcommand{\beq}{\begin{equation}}
\newcommand{\eeq}{\end{equation}}

\begin{document}
 \title{Nonrelativistic scale anomaly, \\ and composite operators with complex scaling dimensions}
 \author{Sergej Moroz}
 \affiliation{Institut f\"{u}r Theoretische Physik Universit\"at Heidelberg Philosophenweg 16, D-69120 Heidelberg, Germany}

\begin{abstract}
It is demonstrated that a nonrelativistic quantum scale anomaly manifests itself in the appearance of composite operators with complex scaling dimensions. In particular, we study nonrelativistic quantum mechanics with an inverse square potential and consider a composite s-wave operator ${\cal O}=\psi\psi$. We analytically compute the scaling dimension of this operator and determine the propagator $\langle0|T{\cal O} {\cal O}^{\dagger} |0\rangle$. The operator ${\cal  O}$ represents an infinite tower of bound states with a geometric energy spectrum. Operators with higher angular momenta are briefly discussed.
\end{abstract}
 
 
 \maketitle

\section{Introduction} \label{intro}
Relativistic conformal field theory is a well-developed subject with numerous applications in statistical physics (systems near a continuous phase transition) and high-energy physics. Recently, Nishida and Son made an important step towards systematic understanding of nonrelativistic conformal field theories (NRCFT) \cite{Nishida, Nishida2}\footnote{for an earlier work see \cite{Mehen}} defined as invariant with respect to the Schr\"odinger symmetry \cite{NiedererHagen}. This symmetry contains the usual Galilei symmetry of nonrelativistic systems extended by a scale and special conformal transformation. The symmetry transformations form a group, called the Schr\"odinger group, which is a direct nonrelativistic analogue of the relativistic conformal group. As a simple example, the free nonrelativistic field theory has the Schr\"odinger symmetry. However, also a number of theories with interactions such as cold fermions at unitarity are believed to be invariant with respect to the Schr\"odinger symmetry, which provides powerful constraints on the correlation functions.

Similar to relativistic conformal field theories the basis of a NRCFT is formed by primary operators \cite{Nishida, Nishida2}. A local primary operator ${\cal O}(t, \vec{x})$ has  a well-defined scaling dimension $\Delta_{{\cal O}}$ and particle number $N_{{\cal O}}$
\beq \label{intro1}
[D,{\cal O}]=i\Delta_{{\cal O}} {\cal O}, \qquad [N,{\cal O}]=N_{{\cal O}}{\cal O},
\eeq
where ${\cal O}\equiv{\cal O}(t=0,\vec{x}=0)$, $D$ and $N$ denote the generators of scale and particle number symmetry, respectively. In addition, the primary operator ${\cal O}$ must commute with the generators of Galilei boost $K_i$ and the special conformal generator $C$
\beq \label{intro2}
[K_i, {\cal O}]=0, \qquad [C, {\cal O}]=0. 
\eeq
The set of descendant operators constructed by subsequent application of time and spatial derivatives on the primary operator ${\cal O}$ form the irreducible representation of the Schr\"odinger algebra \cite{Nishida}. The scaling dimensions of descendants are related in a simple way to the scaling dimension of the parent primary operator ${\cal O}$.

It is known that in some nonrelativistic theories the classical Schr\"odinger symmetry can be broken by a quantum scale anomaly \cite{Camblong}.\footnote{In general, a quantum anomaly means that a classical symmetry is broken at the quantum level due to a regularization and renormalization procedure.} This often leads to discrete scale invariance and a geometric bound state energy spectrum \cite{Efimov, Bedaque} (for a review and source of further references see \cite{BH}). The anomaly also manifests itself in the renormalization group flow of a contact coupling which develops a limit cycle \cite{BH}. Another signature of the nonrelativistic scale anomaly is the appearance of composite operators with complex scaling dimensions. In this paper we concentrate on the latter and discuss some properties of composite operators with complex scaling dimensions in a nonrelativistic quantum theory. In particular, we consider a two-particle problem in the quantum theory with an inverse square potential interaction. In this simple example we analytically compute the propagator of the s-wave composite operator ${\cal O}=\psi\psi$ and examine its structure.

The rest of the paper is organized as follows: In Sec. \ref{qm} we provide a brief introduction to quantum mechanics with an inverse square potential. In Sects. \ref{comp}-\ref{prop} we consider the s-wave composite operator ${\cal O}=\psi\psi$ and calculate its two-point correlation function in momentum space. Sec. \ref{higher} is devoted to a short discussion of composite operators with higher angular momenta. Finally, we close in Sec. \ref{conc} with conclusion and discussion. Technical details are summarized in Appendices.

\section{Quantum mechanics with inverse square potential} \label{qm}
Consider the nonrelativistic quantum theory of identical bose particles interacting through a long-range inverse square potential $V(r)=-\frac{\kappa}{r^2}$ in $D$ spatial dimensions.\footnote{In this work we use field-theoretic (second-quantized) formulation of nonrelativistic quantum theory, which allows to treat different few-body problems in a unified way.} The corresponding nonrelativistic quantum field theory is defined by the microscopic (classical) action
\beq \label{qm1}
\begin{split}
 S[\psi,\psi^{*}]&=\int dt \int d^{D}x \psi^{*}(t, \vec{x})[i\partial_{t}+\frac{\nabla^{2}_{\vec{x}}}{2}]\psi(t, \vec{x}) \\
& -\frac{1}{2}\int dt \int d^{D}x d^{D}y \psi^{*}(t, \vec{x})\psi^{*}(t, \vec{y})V(|\vec{x}-\vec{y}|)\psi(t, \vec{y})\psi(t, \vec{x}).
\end{split}
\eeq
In our convention the particle mass $M_{\psi}$ and the reduced Planck constant $\hbar$ are set to unity. The classical action (\ref{qm1}) is symmetric under the global $U(1)$ internal transformation, associated with the particle number conservation, and is invariant under the Galilean spacetime symmetry group. The interaction parameter $\kappa$ characterizes the strength of the long-range inverse square potential and is positive in the attractive case. Most remarkably, in any spatial dimension the action (\ref{qm1}) is scale invariant because the potential is a homogeneous function of degree $-2$ and has the same scaling as the kinetic part in Eq. (\ref{qm1}).

The two-particle sector of quantum theory with the inverse square potential was studied extensively  \cite{inversesquaregeneral, inversesquarelimitcycle, Camblong,Barford,MS} and by now is well understood. It is a paradigm for nonrelativistic conformal invariance and scale anomaly with a number of physical applications ranging from low-energy atomic to high-energy particle physics (for the list of applications see e.g. \cite{MS}).

The Feynman rules for the particle propagator $iG_{\psi}=\langle 0| T \psi\psi^{\dagger}|0 \rangle$ and the interaction vertex in momentum space are
\beq \label{fig1}
\includegraphics[width=5in]{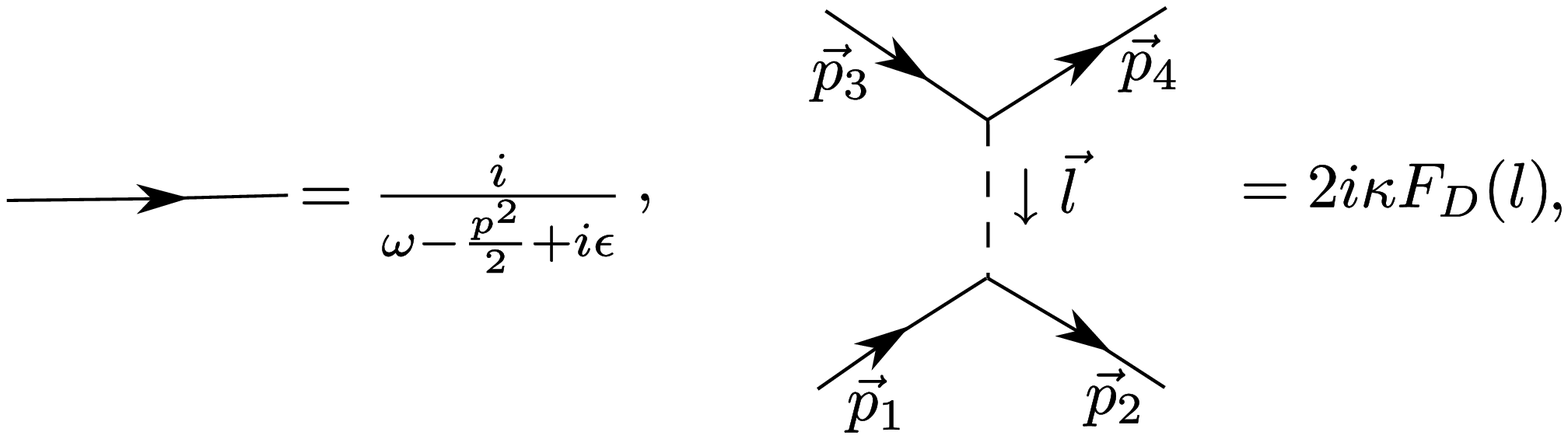}
\eeq
where $i\epsilon$ is an infinitesimal imaginary number, ensuring the retarded causal structure of the propagator. $\vec{l}=\vec{p}_2-\vec{p}_1=\vec{p}_3-\vec{p}_4$ is the spatial momentum transfer during a collision ($l=|\vec{l}|$), and $F_D(l)$ denotes the Fourier transformation of the inverse square potential which in $D>2$ reads
\beq \label{qm2}
F_{D}(l)=\int d^{D}x \frac{1}{x^2}\exp[-i\vec{l}\cdot\vec{x}]=\frac{(4\pi)^{D/2}\Gamma(D/2-1)l^{2-D}}{4}.
\eeq
In the rest of this work we restrict our attention to spatial dimension $D>2$.
\section{Composite operator ${\cal O}=\psi\psi$} \label{comp}
Consider a local two-particle s-wave operator ${\cal O}(t,\vec{x})=\psi\psi(t,\vec{x})$ which annihilates two identical bose particles at the spacetime point $(t,\vec{x})$. As demonstrated in Appendix \ref{app1}, we can assign to the operator ${\cal O}$ a pair of scaling dimensions
\beq \label{comp1}
\Delta_{\pm}=\frac{D+2}{2}\pm\sqrt{\frac{(D-2)^2}{4}-\kappa}.
\eeq
The value of the interaction coupling $\kappa$ determines two qualitatively different regimes:
\begin{itemize}
\item {\bf Undercritical regime} $\kappa<\kappa_{cr}=\frac{(D-2)^2}{4}$: Both $\Delta_{+}$ and $\Delta_{-}$ are real and the action (\ref{qm1}) defines two conformal quantum field theories $CFT_+$ and $CFT_{-}$. More precisely, the quantum field theory defined by the classical action (\ref{qm1}) flows towards $CFT_{+}$ in the infrared and towards $CFT_-$ in the ultraviolet in the renormalization group sense. To obtain the action of $CFT_{+}$ ($CFT_-$),  we must add to the inverse square potential a delta function localized near the origin
\beq \label{comp2}
V(r)\to -\frac{\kappa}{r^2}-\frac{\lambda}{\epsilon} \delta(r-\epsilon)
\eeq
and fine-tune the value of the dimensionless contact coupling $\lambda$ \cite{Kaplan} to the infrared (ultraviolet) fixed point. The operator ${\cal O}$ is a nonrelativistic two-particle primary operator because it is a local product of two primary operators $\psi$.
\item {\bf Overcritical regime} $\kappa>\kappa_{cr}$: The quantum field theory ceases to be conformal and renormalization group evolution of the coupling $\lambda$ in Eq. \eqref{comp2} develops a limit cycle \cite{inversesquarelimitcycle, Camblong,Barford,MS}. $\Delta_{\pm}$ in Eq. \eqref{comp1} becomes complex and form a conjugate pair. Due to the loss of conformality $\Delta_{\pm}$ do not fulfill the definition for scaling dimension as given in Eq. \eqref{intro1}. However, we will still use the term scaling dimensions for $\Delta_{\pm}$ even in the overcritical regime. This is motivated by observation made in \cite{MS} that one can formally extend parameter space of the contact coupling $\lambda$ to the complex plane. In this extension two complex fixed points emerge in the overcritical regime. At these fixed points the operator ${\cal O}$ has scaling dimensions $\Delta_{\pm}$ in the sense of Eq. \eqref{intro1}.
\end{itemize}

We now calculate the two-point function (two-particle propagator)
\beq \label{comp3}
iG_{{\cal O}}(t_2, \vec{x}_2; t_1, \vec{x}_1)=\langle 0|T {\cal O}(t_2, \vec{x}_2) {\cal O}^{\dagger}(t_1, \vec{x}_1) |0\rangle
\eeq
in the overcritical regime $\kappa>\kappa_{cr}$.
Intuitively, the two-point function is proportional to the probability of creating two identical particles at the spacetime point $(t_1, \vec{x}_1)$ and subsequent destruction of the pair at the distinct point $(t_2, \vec{x}_2)$.

\begin{figure}[t]
\begin{center}
\includegraphics[width=4.0in]{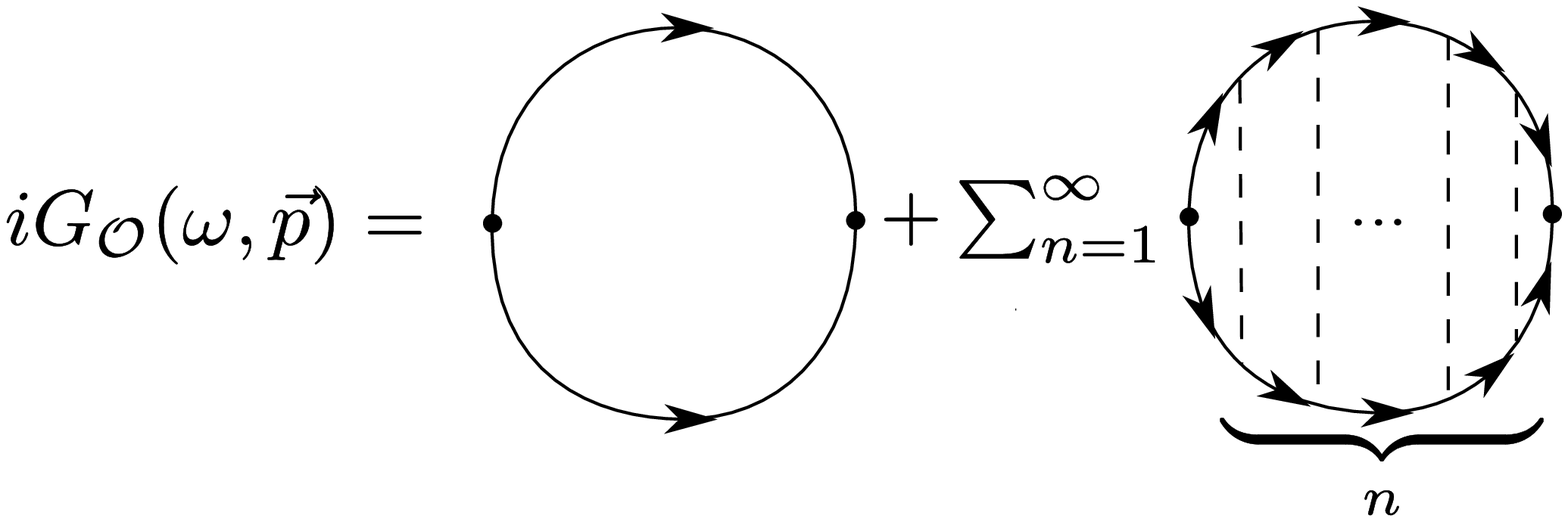}
\end{center}
\vskip -0.7cm \caption{The two-particle propagator $iG_{{\cal O}}(\omega, \vec{p})$ illustrated as a sum of Feynman diagrams. The full circles denote the composite operator insertions.}
\label{fig2}
\end{figure}

The translational invariance of the action (\ref{qm1}) allows us to transform $iG_{{\cal O}}$ to the momentum space
\beq \label{comp4}
iG_{{\cal O}}(\omega, \vec{p})=\int dt d^{D}x e^{i\omega t-i\vec{p}\cdot \vec{x}}iG_{{\cal O}}(t, \vec{x}; 0, \vec{0}).
\eeq
The diagramatic expression of the propagator $iG_{{\cal O}}(\omega, \vec{p})$ is depicted in Fig. \ref{fig2}. The first diagram in Fig. \ref{fig2} corresponds to a free propagation of two particles, while the remaining sum of diagrams takes into account the influence of the long-range interaction.

At this point it must be stressed that few-body problems in nonrelativistic quantum field theory are relatively simple because only particles (but no antiparticles) exist as excitations of the nonrelativistic vacuum.\footnote{From this perspective, many-body problems are more difficult due to a more complicated vacuum state which allows the presence of hole excitations.} This remarkable fact and the particle number conservation lead to the important diagramatic simplification: \emph{Any loop with line arrows pointing in the same direction vanishes.} Mathematically, this can be demonstrated with the help of the residue theorem for the frequency loop integration. For this reason the above-mentioned loops which involve both particles and antiparticles vanish.

The diagramatic simplification has a number of important consequences:
\begin{itemize}
\item The particle propagator is not renormalized and coincides with the bare propagator depicted in Eq. (\ref{fig1}). For this reason there are only bare propagators in Fig. \ref{fig2}.
\item The long-range interaction vertex is not renormalized (i.e. there is no screening). That is why there are only bare vertices in Fig. \ref{fig2}.
\item There are no ``crossed'' interaction lines in Fig. \ref{fig2}.
\end{itemize}

We note that due to the Galilean symmetry of the action (\ref{qm1}), the two-particle propagator $iG_{{\cal O}}(\omega, \vec{p})$ should be a function of only the Galilean invariant combination $\omega-\vec{p}^2/4$. Hence, it is advantageous to switch to the center-of-mass frame (i.e. set $\vec{p}=0$) first and calculate $iG_{{\cal O}}(\omega)\equiv iG_{{\cal O}}(\omega, \vec{p}=0)$. In the very end we can recover momentum dependence by the substitution $iG_{{\cal O}}(\omega)\to iG_{{\cal O}}(\omega-\vec{p}^2/4)$.

\begin{figure}[t]
\begin{center}
\includegraphics[width=6.0in]{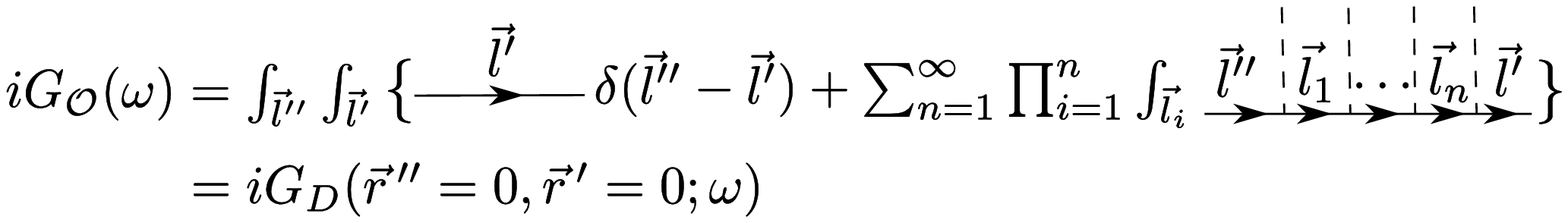}
\end{center}
\vskip -0.7cm \caption{The relation between the two-particle propagator $iG_{{\cal O}}(\omega)$ and the energy Green function $G_{D}(\vec{r}\,'',\vec{r}\,';\omega)$. Full lines with an arrow denote here the bare propagator of a particle of the reduced mass $M_{red}=\frac{1}{2}$. }
\label{fig3}
\end{figure}

Separation of relative and center-of-mass motion in the two-body problem allows us to relate the two-particle propagator $iG_{{\cal O}}(\omega)$ to the energy Green function $G_{D}(\vec{r}\,'',\vec{r}\,';\omega)$ of a single particle with the reduced mass $M_{red}=\frac{M_{\psi}}{2}=\frac{1}{2}$ in the external inverse square potential. To clarify this point, consider the first diagram in Fig. \ref{fig2}, which we shall call $iG_{{\cal O}}^{0}(\omega)$
\beq \label{comp5}
\begin{split}
iG_{{\cal O}}^{0}(\omega)&=\int_{\vec{l}}\int \frac{dl_0}{2\pi}\frac{i}{l_0-\frac{l^2}{2}+i\epsilon}\frac{i}{\omega-l_0-\frac{l^2}{2}+i\epsilon}
 =\int_{\vec{l}}\int_{\vec{l}\,'}\frac{i}{\omega-l^2+i\epsilon} \delta(\vec{l}-\vec{l}\,') \\
&=i\langle \vec{r}\,''=0|(\omega-\hat{p}^2+i\epsilon)^{-1}|\vec{r}\,'=0\rangle=iG_{D}^{0}(\vec{r}\,''=0,\vec{r}\,'=0;\omega),
\end{split}
\eeq
where $\int_{\vec{l}}=\int\frac{d^D l}{(2\pi)^D}$. In the first line we performed frequency loop integration. In the last line the definition of the energy Green function $G_{D}^{0}(\vec{r}\,'',\vec{r}\,';\omega)$ evaluated at the origin $\vec{r}\,''=\vec{r}\,'=0$ was recognized. In the same way, we can perform frequency loop integrations in the diagrams with interaction vertices in Fig. \ref{fig2} and find that
\beq \label{comp6}
iG_{{\cal O}}(\omega)=i\langle \vec{r}\,''=0|(\omega-\hat{H}+i\epsilon)^{-1}|\vec{r}\,'=0\rangle=iG_{D}(\vec{r}\,''=0,\vec{r}\,'=0;\omega),
\eeq 
where $\hat{H}=\hat{p}^2-\frac{\kappa}{\hat{r}^2}$. In terms of Feynman diagrams the last relation can be expressed as depicted in Fig. \ref{fig3}.
The energy Green function $G_{D}(\vec{r}\,'',\vec{r}\,';\omega)$ for the inverse square potential problem was calculated in \cite{Camblong:2001br}, and a modified derivation is presented in the next section.
\section{Energy Green function $G_{D}(\vec{r}\, '',\vec{r}\, ';\omega)$} \label{app2}
The energy Green function in position representation in $D$ spatial dimensions is given by
\beq \label{b1}
G_{D}(\vec{r}\, '',\vec{r}\, ';\omega)=\langle \vec{r}\,''|(\omega-\hat{H}+i\epsilon)^{-1}|\vec{r}\,'\rangle,
\eeq
where $\hat{H}=\hat{p}^2+V(\hat{\vec{r}})$ is a single particle Hamiltonian. The definition (\ref{b1}) leads to the following inhomogeneous differential equation
\beq \label{b2}
\left[\nabla^{2}_{\vec{r}\,''}+\omega-V(\vec{r}\,'')\right]G_{D}(\vec{r}\, '',\vec{r}\, ';\omega)=\delta(\vec{r}\,''-\vec{r}\,').
\eeq
For a central potential Eq. (\ref{b2}) can be solved by separation of variables. In particular, the energy Green function can be expanded in D-dimensional partial waves
\beq \label{b3}
G_{D}(\vec{r}\, '',\vec{r}\, ';\omega)=(r\,'' r\,')^{-\frac{D-1}{2}}\sum_{l=0}^{\infty}\sum_{m=1}^{d_l}Y_{lm}(\Omega\,'')Y_{lm}^{*}(\Omega\,')G_{l}(r\,'',r\,';\omega),
\eeq
where $Y_{lm}(\Omega)$ denote D-dimensional spherical harmonics and $d_l=\frac{(2l+D-2)(l+D-3)!}{l!(D-2)!}$.\footnote{We use the notation from \cite{Camblong:2001br} for the labelling of spherical harmonics in $D$ dimensions.} Substitution of Eq. (\ref{b3}) in Eq. (\ref{b2}) gives the differential equation for the radial energy Green function
\beq \label{b4}
\left[ \frac{d^2}{dr\,''^{2}}+\omega-V(r\,'')-\frac{(l+\nu)^2-\frac{1}{4}}{r\,''^{2}}\right]G_{l}(r\,'',r\,';\omega)=\delta(r\,''-r\,')
\eeq
with $\nu=\frac{D-2}{2}$. Solution of this equation is now a textbook problem (see e.g. \cite{FMRT}). The radial energy Green function can be expressed as
\beq \label{b5}
G_{l}(r\,'',r\,';\omega)=\frac{u_{l}^{(<)}(r_{<})u_{l}^{(>)}(r_{>})}{W\{u_{l}^{(<)},u_{l}^{(>)}\}(r\,')}
\eeq
with $r_{<}$ (resp. $r_{>}$)  denoting the smaller (resp. larger) one of $r\,'$ and $r\,''$. $W\{u_{l}^{(<)},u_{l}^{(>)}\}$  represents the Wronskian of $u_{l}^{<}$ and $u_{l}^{>}$, i.e.
\beq \label{b6}
W\{u_{l}^{(<)},u_{l}^{(>)}\}=u_{l}^{(<)}u_{l}^{(>)}\,'-u_{l}^{(<)}\,' u_{l}^{(>)}.
\eeq
The function $u_{l}^{(<)}$ (resp. $u_{l}^{(>)}$) solves the homogeneous differential equation
\beq \label{b7}
\left[ \frac{d^2}{dr^{2}}+\omega-V(r)-\frac{(l+\nu)^2-\frac{1}{4}}{r^{2}}\right]u_l(r)=0
\eeq
with a regular boundary condition at $r=0$ (resp. $r=\infty$).

Consider now the inverse square potential $V(r)=-\frac{\kappa}{r^2}$. Depending on the value of the coupling strength $\kappa$, the solution of Eq. (\ref{b7}) has two qualitatively different branches. We analyze these two regimes separately.

\subsection{Undercritical regime $\kappa<(l+\nu)^2$}
Two linearly independent solutions of Eq. (\ref{b7}) are
\beq \label{b8}
u_l(r)=\{\sqrt{r}I_{s_l}(\zeta r), \sqrt{r}K_{s_l}(\zeta r) \}
\eeq
with $\zeta=\sqrt{-\omega-i\epsilon}$ and $s_l=\sqrt{(l+\nu)^2-\kappa}$. $I_{s_l}$ and $K_{s_l}$ denote modified Bessel functions. We find
\beq \label{b9}
\begin{split}
& u_l^{(>)}(r)\sim \sqrt{r}K_{s_l}(\zeta r), \qquad u_l^{(>)}(r)\stackrel{r \rightarrow \infty}{\longrightarrow} 0, \\
& u_l^{(<)}(r)\sim \sqrt{r}I_{s_l}(\zeta r), \qquad u_l^{(<)}(r)\stackrel{r \rightarrow 0}{\longrightarrow} 0.
\end{split}
\eeq
Substitution of Eq. (\ref{b9}) in Eq. (\ref{b5}) yields the radial Green function
\beq \label{b10}
G_{l}(r\,'',r\,';\omega)=-\sqrt{r\,'' r\,'}I_{s_l}(\zeta r_{<})K_{s_l}(\zeta r_{>}).
\eeq

\subsection{Overcritical regime $\kappa>(l+\nu)^2$}
In this case two linearly independent solutions of Eq. (\ref{b7}) are found to be
\beq \label{b11}
u_l(r)=\{\sqrt{r}I_{i\theta_l}(\zeta r), \sqrt{r}K_{i\theta_l}(\zeta r) \}
\eeq
with $\theta=\sqrt{\kappa-(l+\nu)^2}$. It is straightforward to determine $u^{(>)}_l$, which is
\beq \label{b12}
u_l^{(>)}(r)\sim \sqrt{r}K_{i\theta_l}(\zeta r), \qquad u_l^{(>)}(r)\stackrel{r \rightarrow \infty}{\longrightarrow} 0.
\eeq
On the other hand, determination of $u_l^{(<)}(r)$ turns out to be more subtle. It is easily demonstrated that any linear combination of two independent solutions (\ref{b11}) approaches zero as $r\to 0$, and thus satisfies the regular boundary condition. Mathematically, the problem originates from the singular behavior of the inverse square potential at $r=0$. In the overcritical regime the potential is too singular and must be regularized. Regularization, introduced in \cite{Camblong:2001br}, imposes the boundary condition at some small but finite $r=a$. Here, following \cite{Barford}, we use a different but equivalent regularization procedure which consists in fixing a phase angle of $u_l^{(<)}(r)$ as $r\to 0$. Specifically, we take the function $u_l^{(<)}(r)$ to be
\beq \label{b13}
u_l^{(<)}(r)\sim \sqrt{r}\left[e^{i\eta(\zeta)}I_{i\theta_l}(\zeta r)+e^{-i\eta(\zeta)}I_{-i\theta_l}(\zeta r) \right]
\eeq
with $\eta(\zeta)=-\theta_l \ln\left(\frac{\zeta}{\zeta^*_l} \right)$. Here we introduced a momentum scale $\zeta^{*}_l$ that determines the phase shift of $u_l^{(<)}(r)$ near the origin
\beq \label{b14}
u_l^{(<)}(r)\sim \sqrt{r}\cos(\theta_l \ln r +\beta+\underbrace{\theta_l \ln \zeta^{*}_l}_{\delta_l\phi}),
\eeq
where $\beta=\text{arg}\left(\frac{2^{-i\theta_l}}{\Gamma(1+i\theta_l)} \right)$. We emphasize that the phase shift $\delta_l\phi$ is a physical parameter that fixes $u_l^{(<)}(r)$ in the unique way (up to normalization). Substitution of Eqs. (\ref{b12}, \ref{b14}) in Eq. (\ref{b5}) yields the radial energy Green function
\beq \label{b15}
G_{l}(r\,'',r\,';\omega)=-\frac{\sqrt{r\,'' r\,'}\left[ e^{i\eta(\zeta)}I_{i\theta_l}(\zeta r_{<})+e^{-i\eta(\zeta)}I_{-i\theta_l}(\zeta r_{<}) \right] K_{i\theta_l}(\zeta r_{>})}{2\cos \left(\theta_l \ln (\zeta/\zeta^*_l) \right)}.
\eeq 
\section{Two-particle propagator $iG_{{\cal O}}$} \label{prop}
We are now in the position to finish the calculation of the two-particle propagator $iG_{{\cal O}}(\omega)$ in the overcritical (anamalous) regime. To this end,  according to Eq. (\ref{comp6}) one must evaluate the energy Green function $G_{D}(\vec{r}\, '',\vec{r}\, ';\omega)$ at $\vec{r}\, ''=\vec{r}\, '=0$. Inspection of Eqs. (\ref{b3},\ref{b15}) reveals, however, that for $D>2$ this leads to a divergent result. This is not a surprise because the operator ${\cal O}$ is a local product of two elementary bose operators $\psi$. Composite operators usually lead to additional divergences in quantum field theory which must be renormalized separately. We will deal with this problem in Appendices \ref{appb} and \ref{appc}. The propagator of the bare operator ${\cal O}$ reads
\beq \label{prop1}
iG_{{\cal O}}(\omega)=\langle 0|T {\cal O} {\cal O}^{\dagger} |0\rangle=i\lim_{r\to 0}\int \frac{d\Omega\,'}{S^{D-1}} \int \frac{d\Omega\,''}{S^{D-1}} G_{D}(\vec{r}\, '',\vec{r}\, ';\omega)\Big|_{r\,''=r\,'=r},
\eeq 
where $S^{D-1}=2\frac{\pi^{D/2}}{\Gamma(D/2)}$. In the last expression we introduced two angular integration averages that should be performed before the radial limit. This ensures the final result to be independent of the directions of $\vec{r}\,'$ and $\vec{r}\,''$. We now substitute the partial wave expansion (\ref{b3}) in Eq. (\ref{prop1}) and obtain
\beq \label{prop2}
iG_{{\cal O}}(\omega)=-i\lim_{r\to 0} r^{2-D}\frac{\left[ e^{i\eta(\zeta)}I_{i\theta_0}(\zeta r)+e^{-i\eta(\zeta)}I_{-i\theta_0}(\zeta r) \right] K_{i\theta_0}(\zeta r)}{2\cos \left(\theta_0 \ln (\zeta/\zeta^*_0) \right)},
\eeq
where only the s-wave ($l=0$) survives the angular integrations. Finally, we perform the limit $r\to0$ and find
\beq \label{prop3}
iG^{ren}_{{\cal O}}(\omega)=i\tan\left(\theta_0 \ln\frac{\sqrt{-\omega-i\epsilon}}{\zeta^{*}_0} \right),
\eeq
where computational details and the explicit definition of the renormalized two-particle propagator $iG^{ren}_{{\cal O}}(\omega)$ can be found in Appendix \ref{appb}. For the sake of completeness we compute  $i G^{ren}_{\cal O}(\omega)$ in the undercritical (conformal) regime in Appendix \ref{appc}.

Our final result (\ref{prop3}) is remarkably simple and has the following properties:
\begin{itemize}
\item For negative energies $\omega<0$ the two-particle propagator $iG^{ren}_{{\cal O}}(\omega)$ has an infinite number of simple pole divergences at
\beq \label{prop4}
\omega_n=-\zeta^{*2}_0\exp\left(-\frac{2\pi n}{\theta_0}+\frac{\pi}{\theta_0} \right), \quad n\in\mathbb{Z}.
\eeq
Hence, the composite field ${\cal O}^{\dagger}$ represents infinitely many different stable particles (two-particle s-wave bound states). The energy spectrum has an accumulation point at $\omega\to 0$ as $n\to\infty$. It exhibits geometric behavior
\beq \label{prop5}
\frac{\omega_{n+1}}{\omega_{n}}=\exp\left(-\frac{2\pi}{\theta_0}\right).
\eeq
For positive energies $\omega>0$ the two-point correlation function $iG^{ren}_{{\cal O}}(\omega)$ has a branch cut with a branch point at $\omega=0$.
\item The regularization momentum scale $\zeta^*_0$ introduced in Sec. \ref{app2} is a physical parameter of the theory. It determines the value of the reference (for example $n=0$) two-particle bound state energy
\beq \label{prop6}
\omega_0=-\zeta^{*2}_0 \exp\left(\frac{\pi}{\theta_0} \right).
\eeq
\item The propagator $iG^{ren}_{{\cal O}}(\omega)$ is invariant under the discrete scaling symmetry
\beq \label{prop7}
\zeta^*_0\to\zeta^*_0 \exp\left(\frac{\pi m}{\theta_0} \right), \quad m\in\mathbb{Z}.
\eeq
This transformation maps $\omega_n\to\omega_{n+m}$, but does not alter the energy spectrum. Physically inequivalent $\zeta^*_0$ may thus be chosen to lay within the interval $\zeta^{*}_0\in [1, \exp\left(\frac{\pi}{\theta_0} \right))$. This supports our observation in Sec. \ref{app2} that $\zeta^{*}_0$ determines the phase shift angle $\delta_0\phi=\theta_0\ln\zeta^*_0$ of the wave function near the origin. We conclude that physically inequivalent phase shifts $\delta_0\phi$ span the interval $[0,\pi)$.
\item Finally, as advocated in Sec. \ref{comp}, the Galilean symmetry allows us to recover the momentum dependence of the two-particle propagator 
\beq \label{prop8}
iG^{ren}_{{\cal O}}(\omega, \vec{p})=i\tan\left(\theta_0 \ln\frac{\sqrt{-\omega+\frac{\vec{p}^2}{4}-i\epsilon}}{\zeta^{*}_0} \right).
\eeq
The position space representation of the propagator is derived in Appendix \ref{appd}.
\end{itemize}

\section{Composite operators ${\cal O}^{(l)}$ with higher angular momentum} \label{higher}
The energy Green function \eqref{b1} contains also information about the two-particle composite operators ${\cal O}^{(l)}$ which carry the higher angular momentum $l$. In the conformal regime, these operators are nonrelativistic primaries of ref. \cite{Nishida}. The p-wave and d-wave operators in a theory with a one-component bose field $\psi$ are explicitly given by
\beq \label{h1a}
{\cal O}^{(l=1)}_{i}=\nabla_i \psi\psi-\psi \nabla_i\psi,
\eeq
\beq \label{h1b}
 {\cal O}^{(l=2)}_{ij}=\left[\psi\nabla_i\nabla_j\psi-\nabla_i\psi\nabla_j\psi\right]-\frac{\delta_{ij}}{D}\left[\psi\Delta\psi-\nabla_k \psi \nabla_k \psi \right] 
\eeq
Up to normalization these operators are fixed by two requirements. First, they must be constructed from only two elementary fields $\psi$ and spatial gradients. Second, they are nonrelativistic primaries of ref. \cite{Nishida} and satisfy Eqs. (\ref{intro1}, \ref{intro2}). We expect that in a similar fashion the two-particle primaries with $l>2$ can be constructed. It is clear from Eq. \eqref{h1a} that in the bosonic theory \eqref{qm1} the p-wave operator ${\cal O}^{(l=1)}_{i}$ vanishes, as well as all primary operators with odd angular momentum. The d-wave operator ${\cal O}^{(l=2)}_{ij}$ is a symmetric, traceless tensor.\footnote{In order to make the tensor traceless, one subtracts in Eq. \eqref{h1b} the s-wave part ${\cal O}_s=\psi\Delta\psi-\nabla_k \psi \nabla_k \psi$. It is instructive to express this operator as ${\cal O}_s=-\frac{1}{2}\nabla_k \nabla_k {\cal O}-2i \partial_t {\cal O}+4 \psi\left(i\partial_t+\frac{\Delta}{2} \right)\psi$, where the first and the second terms are the descendants of the primary ${\cal O}$ and the third term is a primary operator.}

The scaling dimensions of the composites ${\cal O}^{(l)}$ were computed in Appendix \ref{app1} and are given by
\beq \label{h2}
\Delta^{(l)}_{\pm}=\frac{D+2}{2}\pm \sqrt{\left(l+\frac{D-2}{2}\right)^2-\kappa}.
\eeq
They become complex for $\kappa>\left(l+\frac{D-2}{2} \right)^2$.

The two-particle propagator $iG_{{\cal O}^{(l)}}=\langle 0|T {\cal O}^{(l)} {\cal O}^{(l)\dagger} |0\rangle$ can be extracted from the $l^{\text{th}}$ partial wave of the energy Green function, i.e. it is encoded in the radial Green function $G_l(r'',r';\omega)$. Following closely the steps from Sec. \ref{prop}, in the overcritical regime we obtain for the renormalized propagator
\beq \label{h3}
iG_{{\cal O}^{(l)}}^{ren}(\omega,\vec{p})=i\tan\left(\theta_l \ln\frac{\sqrt{-\omega+\frac{\vec{p}^2}{4}-i\epsilon}}{\zeta^{*}_l} \right)T_{i_1\dots i_l,j_1\dots j_l}(\vec{p}),
\eeq
where $T_{i_1\dots i_l,j_1\dots j_l}$ is a tensor symmetric and traceless in the indices $i_1\dots i_l$ and $j_i\dots j_l$. This tensor does not depend on the energy $\omega$, and thus we can extract the energies of the two-particle bound states with angular momentum $l$ from Eq. \eqref{h3}. Similar to the s-wave bound states, for even angular momenta $l$ the energy spectrum exhibits geometric behavior
\beq \label{h4}
\frac{\omega_{n+1}}{\omega_{n}}=\exp\left(-\frac{2\pi}{\theta_l}\right).
\eeq
\section{Conclusions} \label{conc}
In this work we examined composite operators in a nonrelativistic quantum field theory. In general, the presence of such operators is a signature of the quantum scale anomaly. As a concrete example, we considered quantum mechanics with the classically scale invariant inverse square potential and studied the two-particle local scalar composite operator ${\cal O}=\psi\psi$. We determined the complex scaling dimension of this operator and analytically calculated the two-particle propagator $\langle0| T {\cal O} {\cal O}^{\dagger} |0\rangle$. In the nonconformal (anomalous) regime the operator ${\cal O}$ represents an infinite tower of s-wave two-particle bound states which form a geometric energy spectrum.

For simplicity, in this work we considered the specific two-particle inverse square potential problem, which is the paradigmatic example of a nonrelativistic theory with the quantum scale anomaly. Nevertheless, we expect that our simple result \eqref{prop8} for the two-point correlator $\langle0| T {\cal O} {\cal O}^{\dagger} |0\rangle$ is universal and applies to other more complicated nonrelativistic theories containing composite operators with complex scaling dimensions. One prominent example is a system of three identical bosons interacting  through a contact potential tuned to the unitarity point. Due to the Efimov effect \cite{Efimov}, the local atom-dimer composite operator ${\cal O}=\psi\phi$ acquires the complex scaling dimensions
\beq \label{conc1}
\Delta_{\pm}=\frac{5}{2}\pm i s_{0},
\eeq
where the Efimov parameter $s_0\approx 1.00624$. A similar system that exhibits the quantum scale anomaly in the three-particle sector is the two-component fermionic system with unequal masses \cite{Efimov}. Another interesting example is the one-dimensional model with a four-body resonant interaction \cite{Nishida3} where the scaling dimension of the five-boson composite operator is complex, too.

Recently we used the method of the nonrelativistic AdS/CFT correspondence to study operators with complex scaling dimensions \cite{Moroz}. Using the standard AdS/CFT technique, we calculated the two-point correlation function $\langle 0| {\cal O} {\cal O}^{\dagger} |0\rangle$. The result has the same functional form as found in this work in Eq. \eqref{prop8}. However, while Eq. \eqref{prop8} depends on the physical momentum parameter $\zeta_0^*$, the holographic result of \cite{Moroz} depends on the value of the UV momentum cutoff. The latter is unphysical, and we expect that the cutoff dependence in the AdS/CFT calculation can be eliminated by inclusion of a proper boundary counterterms \cite{progress}.

\emph{Acknowledgments} -- It is our pleasure to specially acknowledge discussions with Mike Birse and Yusuke Nishida. We are thankful to H.~Hammer, D.~Kaplan and D.~Son  for advice. We thank the Institute for Nuclear Theory at the University of Washington where part of this work was completed for its hospitality. We thank Mike Birse and Diana Morozova for critical remarks on the manuscript. The author is supported by KTF.
\appendix
\section{Scaling dimension of composite operator ${\cal O}=\psi\psi$} \label{app1}
An elegant way to calculate the scaling dimensions of the composite two-particle operator ${\cal O}=\psi\psi$ (and more generally of the two-particle primary operators ${\cal O}^{(l)}$ carrying the angular momentum $l$) is to employ the operator/state correspondence \cite{Tan, Castin, Nishida}. To this end one considers two particles interacting through the inverse square potential confined in a harmonic trap. The total Hamiltonian $H$ of this system can be separated into the center-of-mass and relative parts
\beq \label{a1}
\begin{split}
& H=H_{R}+H_{r}, \\
& H_{R}=-\frac{1}{4}\nabla_{\vec{R}}^{2}+\omega^2 R^2,  \qquad E_{R}^{0}=\omega\frac{D}{2}, \\
& H_{r}=-\nabla_{\vec{r}}^{2}-\frac{\kappa}{r^2}+\frac{\omega^2 r^2}{4}, \qquad E_{r,l}^{\pm}=\omega\left(1\pm \sqrt{\left(l+\frac{(D-2)}{2}\right)^2-\kappa} \right),
\end{split}
\eeq
where $\vec{R}$ and $\vec{r}$ are the center-of-mass and relative coordinates. In addition, $E_{R}^{0}$ denotes the ground state energy of $H_R$, and $E_{r,l}^{\pm}$ stands for the lowest energy of the Hamiltonian $H_r$ in the subspace of states with the angular momentum $l$.\footnote{The relative Hamiltonian $H_r$ defines a quantum mechanical problem of a particle in the combined inverse square and harmonic potential, also known as the Calogero problem. The energy spectrum is formally given by two equidistant towers built on top of two ``lowest state energies'' $E_{r,l}^{+}$ and $E_{r,l}^{-}$ \cite{Calogero}. Physically, there are two proper choices of the near-origin boundary condition of the wave function that distinguish between $+$ and $-$ branches of the energy spectrum. The choice $E_{r,l}^{+}$ ($E_{r,l}^{-}$) corresponds to $CFT_+$ ($CFT_-$) of Sec. \ref{comp}.} According to the operator/state correspondence, the scaling dimension of the composite primary ${\cal O}^{(l)}$, carrying the angular momentum $l$, coincides with the lowest energy $E_l^{0}$ of the total Hamiltonian $H$, expressed in the units of the trapping frequency $\omega$
\beq \label{a2}
\Delta^{(l)}_{\pm}=\frac{E_l^{0}}{\omega}=\frac{D+2}{2}\pm \sqrt{\left(l+\frac{D-2}{2}\right)^2-\kappa}.
\eeq
Note that the operator ${\cal O}^{(l)}$ composed of two identical bose (fermi) fields vanishes if the angular momentum $l$ is odd (even). This is due to the fact that two identical bosons (fermions) can not be in the quantum state with odd (even) angular momentum. For the s-wave ($l=0$) operator ${\cal O}^{(l=0)}=\psi\psi$ we thus obtain
\beq \label{a2a}
\Delta_{\pm}\equiv\Delta^{(l=0)}_{\pm}=\frac{D+2}{2}\pm \sqrt{\left(\frac{D-2}{2}\right)^2-\kappa}.
\eeq

We must stress that the operator/state correspondence applies only to nonrelativistic conformal field theories, and thus our result (\ref{a2}) holds for $\kappa<\kappa_{cr}=\left(l+ \frac{D-2}{2}\right)^2$, when both $\Delta^{(l)}_{\pm}$ are real. Nevertheless, it turns out that even in the anomalous (nonconformal) regime for $\kappa>\kappa_{cr}$ Eq. (\ref{a2}) leads to the correct scaling dimensions $\Delta^{(l)}_{\pm}$. We illustrate this fact on the example of the s-wave composite operator ${\cal O}(t, \vec{x})$. Following the observation, made in \cite{Nishida}, the proper definition of the composite ${\cal O}(t, \vec{x})$ for any value of $\kappa$ is given by
\beq \label{a3}
{\cal O}(t, \vec{x})=\lim_{\vec{y}\to\vec{x}}|\vec{x}-\vec{y}|^{-\gamma}\psi(t,\vec{x})\psi(t,\vec{y}),
\eeq
where $\gamma$ is a leading near-origin power law exponent of the zero-energy wave function of the relative Hamiltonian $H_r$. It can be determined from the equation $H_r r^\gamma=0$ and reads
\beq \label{a4}
\gamma=1-\frac{D}{2}\pm \sqrt{\left(\frac{D-2}{2}\right)^2-\kappa}
\eeq
The prefactor $|\vec{x}-\vec{y}|^{\gamma}$ in Eq. \eqref{a3} is needed to make matrix elements of the  operator ${\cal O}(t, \vec{x})$ between any two states in the Hilbert space finite. From Eq. \eqref{a3} we can read off the scaling dimension of the operator ${\cal O}$ by a simple counting
\beq \label{a5}
\Delta_{\pm}=2\Delta_{\psi}+\gamma=\frac{D+2}{2}\pm\sqrt{\left(\frac{D-2}{2}\right)^2-\kappa}
\eeq
which is in agreement with Eq. \eqref{a2a}, found from the operator/state correspondence.

\section{Details of calculation of $iG^{ren}_{{\cal O}}(\omega)$} \label{appb}
First, the limit $r\to0$ in Eq. (\ref{prop2}) can be simplified by exploiting the near-origin asymptotics of the modified Bessel functions
\beq \label{appb1}
\begin{split}
& I_{i\theta_0}(x)\stackrel{x \rightarrow 0}{\longrightarrow} b e^{i(\theta_0\ln x+\beta)}, \qquad be^{i\beta}=\frac{2^{-i\theta_0}}{\Gamma(1+i\theta_0)}, \\
& K_{i\theta_0}(x)\stackrel{x \rightarrow 0}{\longrightarrow} a(e^{i(\theta_0\ln x+\alpha)}+e^{-i(\theta_0 \ln x+\alpha)}), \qquad a e^{i\alpha}=2^{-1-i\theta_0}\Gamma(-i\theta_0)
\end{split}
\eeq
with the result
\beq \label{appb2}
\begin{split}
iG_{{\cal O}}(\omega)=&-2iab\lim_{r\to 0}r^{2-D}\frac{\cos(\theta_0\ln\zeta^*_0r+\beta)\cos(\theta_0\ln\zeta r+\alpha)}{\cos\left(\theta_0\ln(\zeta/\zeta^*_0)\right)} \\
=&2iab\lim_{r\to 0}r^{2-D}\cos(\theta_0\ln\zeta^*_0r+\beta)\sin(\theta_0\ln\zeta^*_0r+\alpha)\times \\ &\left[\tan\left(\theta_0 \ln\frac{\zeta}{\zeta^{*}_0} \right)-\cot(\theta_0 \ln \zeta^*_0r+\alpha) \right],
\end{split}
\eeq
In the second line we used the identities $\ln\zeta r=\ln \zeta^*_0 r+\ln(\zeta/\zeta^*_0)$ and $\cos(A+B)=\cos A \cos B- \sin A \sin B$. 
At this point we observe that the overal normalization factor and the second term in the bracket of Eq. \eqref{appb2} are actually ill-defined in the limit $r\to0$. Physically, this problem originates from the renormalization group limit cycle scaling in the overcritical regime. The limit $r\to0$ is intricate and will be performed in two separate steps. First, we subtract the energy-independent second term in the bracket of Eq. \eqref{appb2}. This choice effectively prescribes the initial position on the limit cycle in the UV and thus fixes the energy spectrum (see Eq. \eqref{prop4}). In the second step we perform a multiplicative renormalization of the composite operator ${\cal O}$ by introducing its renormalized version
\beq \label{appb2a}
{\cal O}^{ren}={\cal N} r^{\frac{D-2}{2}} \left[\cos(\theta_0\ln\zeta^*_0r+\beta)\sin(\theta_0\ln\zeta^*_0r+\alpha)\right]^{-1/2}  {\cal O}
\eeq
with ${\cal N}=\left(2ab \right)^{-1/2}$. Here we emphasize that the renormalization function is not a simple power of $r$ because in the overcritical regime the theory is not conformal but undergoes the limit cycle RG flow. After these two renormalization steps we obtain the renormalized two-particle propagator $iG^{ren}_{{\cal O}}(\omega)$
\beq \label{appb3}
iG^{ren}_{{\cal O}}(\omega)=\langle 0|T {\cal O}^{ren} {\cal O}^{\dagger\,ren} |0\rangle=i\tan\left(\theta_0 \ln\frac{\sqrt{-\omega-i\epsilon}}{\zeta^{*}_0} \right),
\eeq 
where we substituted $\zeta=\sqrt{-\omega-i\epsilon}$. Finally, we note that although the renormalization procedure might appear ad hoc at first sight, in general it parallels its undercritical (conformal) counterpart which is described in Appendix \ref{appc}.

\section{Two-particle propagator in undercritical regime} \label{appc}
In the undercritical (conformal) regime we follow directly the steps from Sec \ref{prop} and obtain
\beq \label{appc1}
\begin{split}
iG_{{\cal O}}(\omega)&=\langle 0|T {\cal O} {\cal O}^{\dagger} |0\rangle=i\lim_{r\to 0}\int \frac{d\Omega\,'}{S^{D-1}} \int \frac{d\Omega\,''}{S^{D-1}} G_{D}(\vec{r}\, '',\vec{r}\, ';\omega)\Big|_{r\,''=r\,'=r} \\
&= -i\lim_{r\to 0}r^{2-D}I_{s_0}(\zeta r)K_{s_0}(\zeta r),
\end{split}
\eeq
where in the second line we used Eqs. (\ref{b3}, \ref{b10}). Now by employing the near-origin asymptotic behavior of the Bessel functions \eqref{appb1} for $\theta_0=-i s_0$ and $x=\zeta r$, we arrive at
\beq \label{appc2}
iG_{{\cal O}}(\omega)=-iabe^{i(\alpha+\beta)}\lim_{r\to 0} r^{2-D} \left((\zeta r)^{2s_0}+e^{-2i\alpha} \right).
\eeq
The bare two-particle propagator is divergent for $D>2$ and must be renormalized. This can be achieved in two steps, and we present the procedure for the example of $CFT_{+}$. In the first step we neglect the energy-independent second term in the bracket of Eq. \eqref{appc2}. Physically, this corresponds to probing the large distance infrared physics which is governed by $CFT_{+}$ (see Sec. \ref{comp}). In the second step we perform a multiplicative renormalization by introducing a renormalized operator
\beq \label{appc2aa}
{\cal O}^{ren}={\cal N} r^{\frac{D-2}{2}-s_0} {\cal O}.
\eeq
This is necessary due to the composite nature of the operator ${\cal O}$. Here for simplicity we absorbed the finite constant ${\cal N}=\left(-ab e^{i(\alpha+\beta)} \right)^{-1/2}$ into the definition of the renormalized operator. With this definition we obtain
\beq \label{appc2a}
iG^{ren}_{{\cal O}}(\omega)=\langle 0|T {\cal O}^{ren} {\cal O}^{\dagger\,ren} |0\rangle=i\zeta^{2s_0}=i\left(-\omega-i\epsilon \right)^{s_0}.
\eeq 
Finally, by recovering the momentum dependence, we obtain a cutoff-independent, renormalized two-particle propagator
\beq \label{appc3}
iG^{ren}_{{\cal O}}(\omega, \vec{p})=i\left(-\omega+\frac{p^2}{4}-i\epsilon \right)^{s_0}.
\eeq
It is straightforward to transform to the position space
\beq \label{appc4}
\begin{split}
iG^{ren}_{{\cal O}}(t, \vec{x})&= \int \frac{d\omega}{2\pi} \frac{d^{D}p}{(2\pi)^D} e^{-i\omega t+i\vec{p}\cdot \vec{x}}iG^{ren}_{{\cal O}}(\omega, \vec{p}) \\
&=C_D \theta(t) t^{-\Delta_+}\exp\left(-iN_{\cal O}\frac{\vec{x}^2}{2t} \right)
\end{split}
\eeq
with $C_D=\frac{i^{3s_0+1}}{\Gamma(-s_0)}\left(\frac{-i}{\pi} \right)^{D/2}$, $\Delta_+=\frac{D+2}{2}+s_0$ and $N_{\cal O}=-2$. This is a familiar result for the retarded propagator of an operator ${\cal O}$ with the scaling dimension $\Delta_+$ and the particle number $N_{\cal O}=-2$ in a nonrelativistic conformal field theory. Its functional form (up to the normalization constant $C_D$) is fixed by the Schr\"odinger symmetry \cite{Nishida}.

\section{Two-particle propagator in position space} \label{appd}
In this appendix we attempt to compute the two-particle propagator in the position space. This can be achieved via the inverse Fourier transformation of Eq. (\ref{prop8})
\beq \label{appd1}
iG^{ren}_{{\cal O}}(t, \vec{x})= \int \frac{d\omega}{2\pi} \frac{d^{D}p}{(2\pi)^D} e^{-i\omega t+i\vec{p}\cdot \vec{x}}i\tan\left(\theta_0 \ln\frac{\sqrt{-\omega+\frac{\vec{p}^2}{4}-i\epsilon}}{\zeta^{*}_0} \right).
\eeq
First we perform the angular integration and obtain
\beq \label{appd2}
iG^{ren}_{{\cal O}}(t, \vec{x})=\left(\frac{1}{2\pi}\right)^{D/2}\int_{0}^{\infty}dp p \left(\frac{p}{x} \right)^{D/2-1}J_{D/2-1}(px)\int\frac{d\omega}{2\pi}e^{-i\omega t}i\tan\left(\theta_0 \ln\frac{\sqrt{-\omega+\frac{p^2}{4}-i\epsilon}}{\zeta^{*}_0} \right).
\eeq
Now we introduce a new variable $W\equiv-\omega+\frac{p^2}{4}$ and get
\beq \label{appd3}
iG^{ren}_{{\cal O}}(t, \vec{x})=\left(\frac{1}{2\pi}\right)^{D/2}\int_{0}^{\infty}dp p \left(\frac{p}{x} \right)^{D/2-1}J_{D/2-1}(px) e^{-i\frac{p^2}{4}t} \int \frac{dW}{2\pi} e^{iW t}i\tan\left(\theta_0 \ln\frac{\sqrt{W-i\epsilon}}{\zeta^{*}_0} \right).
\eeq
Finally, the momentum integral can be done analytically with the result
\beq \label{appd4}
iG^{ren}_{{\cal O}}(t, \vec{x})=\left(\frac{-i}{\pi t} \right)^{D/2}\exp\left(-iN_{\cal O}\frac{\vec{x}^2}{2t} \right) {\cal S}(t),
\eeq
where $N_{\cal O}=-2$ and ${\cal S}(t)=\int \frac{dW}{2\pi} e^{iW t}i\tan\left(\theta_0 \ln\frac{\sqrt{W-i\epsilon}}{\zeta^{*}_0} \right)$. We were not able to perform the integral over $W$ explicitly. Based on the dimensional argument ${\cal S}(t)=t^{-1}f\left((\zeta^{*}_0)^2t, \theta_0\right)$, where $f$ is some function of the dimensionless arguments $(\zeta^{*}_0)^2 t$ and $\theta_0$. We checked that the function ${\cal S}(t)$ is not restricted by the Galilean symmetry.


\begin{thebibliography}{99}
\bibitem{Nishida}
  Y.~Nishida and D.~T.~Son,
  Phys.\ Rev.\  D {\bf 76}, 086004 (2007).
\bibitem{Nishida2}
  Y.~Nishida and D.~T.~Son,
  arXiv:1004.3597.
\bibitem{Mehen}
  T.~Mehen, I.~W.~Stewart and M.~B.~Wise,
  Phys.\ Lett.\  B {\bf 474}, 145 (2000).
\bibitem{NiedererHagen}
  U.~Niederer, 
  Helv.\ Phys.\ Acta {\bf 45}, 802 (1972);
  C.~R.~Hagen,
  Phys.\ Rev.\  D {\bf 5}, 377 (1972).
\bibitem{Camblong}
  H.~E.~Camblong and C.~R.~Ordonez,
  Phys.\ Rev.\  D {\bf 68}, 125013 (2003).
\bibitem{Efimov}
   V.~Efimov, 
   Phys.\ Lett.\  \textbf{33B}, 563 (1970); 
   Nucl.\ Phys.\ A\  \textbf{210}, 157 (1973).
\bibitem{Bedaque}
  P.~F.~Bedaque, H.~W.~Hammer and U.~van Kolck,
  Phys.\ Rev.\ Lett.\  {\bf 82}, 463 (1999);
  Nucl.\ Phys.\  A {\bf 646}, 444 (1999).
\bibitem{BH}
     E. Braaten, H.W. Hammer,
     Phys. Rept. \textbf{428}, 259 (2006).

\bibitem{inversesquaregeneral}
  K.~M.~Case,
  Phys.\ Rev.\  {\bf 80}, 797 (1950);
  V.~de Alfaro, S.~Fubini and G.~Furlan,
  Nuovo Cim.\  A {\bf 34}, 569 (1976);
  E.~Kolomeisky and J.~P.~Straley,
  Phys.\ Rev.\  B {\bf 46}, 12664 (1992);
  K.~S.~Gupta and S.~G.~Rajeev,
  Phys.\ Rev.\  D {\bf 48}, 5940 (1993);
  T.~Barford and M.~C.~Birse,
  Phys.\ Rev.\  C {\bf 67}, 064006 (2003).
\bibitem{inversesquarelimitcycle}
S.~R.~Beane, P.~F.~Bedaque, L.~Childress, A.~Kryjevski, J.~McGuire and U.~v.~Kolck,
  Phys.\ Rev.\  A {\bf 64}, 042103 (2001);
M.~Bawin and S.~A.~Coon,
  Phys.\ Rev.\  A {\bf 67}, 042712 (2003);
E.~J.~Mueller and T.~L.~Ho,
  arXiv:cond-mat/0403283;
  E.~Braaten and D.~Phillips,
  Phys.\ Rev.\  A {\bf 70}, 052111 (2004);
  H.~W.~Hammer and B.~G.~Swingle,
  Annals Phys.\  {\bf 321}, 306 (2006);
  H.~W.~Hammer and R.~Higa,
  Eur.\ Phys.\ J.\  A {\bf 37}, 193 (2008).
\bibitem{Barford}
T.~Barford and M.~C.~Birse,
  J.\ Phys.\ A  {\bf 38}, 697 (2005);
 \bibitem{MS}
   S.~Moroz and R.~Schmidt,
   Annals Phys.\  {\bf 325}, 491 (2010).
 \bibitem{Kaplan}
   D.~B.~Kaplan, J.~W.~Lee, D.~T.~Son and M.~A.~Stephanov,
   Phys.\ Rev.\  D {\bf 80}, 125005 (2009).
\bibitem{Camblong:2001br}
  H.~E.~Camblong and C.~R.~Ordonez,
  Mod.\ Phys.\ Lett.\  A {\bf 17}, 817 (2002); Int.\ J.\ Mod.\ Phys.\  A {\bf 19}, 1413 (2004).
\bibitem{FMRT}
  H.~Feshbach and P.~M.~Morse,
  ``Methods of Theoretical Physics I'', McGrow-Hill book company, 1953;
  L.~Rodberg and R.~Thaler,
  ``Introduction to the quantum theory of scattering'', Academic Press, New York, 1967.
\bibitem{Nishida3}
  Y.~Nishida and D.~T.~Son,
  arXiv:0908.2159.
\bibitem{Moroz}
  S.~Moroz,
  Phys.\ Rev.\  D {\bf 81}, 066002 (2010).
\bibitem{progress}
  S.~Moroz,
  work in progress.

\bibitem{Tan}
S.~Tan,
arXiv:cond-mat/0412764.
\bibitem{Castin}
F.~Werner and Y.~Castin,
Phys.\ Rev.\ Lett. {\bf 97}, 150401 (2006); Phys.\ Rev.\ A {\bf 74}, 053604 (2006).
\bibitem{Calogero}
   F.~Calogero,
   J.\ Math.\ Phys.\  {\bf 10}, 2191 (1969).
 





\end{thebibliography}
\end{document}